\documentclass[onecolumn]{revtex4}

\usepackage{graphicx}
\usepackage{dcolumn}
\usepackage{bm}

\input epsf

\begin{document}

\title{\Large The generalized second law of thermodynamics and the nature of the Entropy
Function}

\author{\bf~Subenoy~Chakraborty\footnote{schakraborty@math.jdvu.ac.in}, Nairwita~Mazumder\footnote{nairwita15@gmail.com},
Ritabrata~Biswas\footnote{biswas.ritabrata@gmail.com}.}

\affiliation{$^1$Department of Mathematics,~Jadavpur
University,~Kolkata-32, India.}

\date{\today}

\begin{abstract}
In black hole physics, the second law of thermodynamics is
generally valid whether the black hole is a static or a non-static
one. Considering  the universe as a thermodynamical system  the
second law of black hole dynamics extends to the non-negativity of
the sum of the entropy of the matter and the horizon, known as
generalized second law of thermodynamics(GSLT). Here, we have
assumed the universe to be bounded by the event-horizon where
Bekenstein entropy-area relation and Hawking-temperature are not
applicable. Thus considering entropy to be an arbitrary function
of the area of the event-horizon, we have tried to find the nature
of the entropy-function for the validity of the GSLT both in
quintessence-era and in phantom-era. Finally, some graphical
representation of the entropy-function has been presented.
\end{abstract}


\maketitle

In black hole physics, semi-classical description shows that just
like a black body , black hole emits thermal radiation (known as
Hawking radiation) and it completes the missing link between a
black hole and a thermodynamical system. The temperature (known as
the Hawking temperature) and the entropy (known as Bekenstein
entropy) are proportional to the surface gravity at the horizon
and area of the horizon \cite{Hawking, Bekenstein} respectively
(i.e. related to the geometry of the horizon). Also this
temperature , entropy and mass of the black hole satisfy the first
law of thermodynamics \cite{Bardeen}. As a result , physicists
start speculating about the relationship between the black hole
thermodynamics and Einstein's field equations (describing the
geometry of space time). It is Jacobson \cite{Jacobson} who first
derived Einstein field equations from the first law of
thermodynamics: $\delta Q=TdS $ for all local Rindler causal
horizons with $\delta Q$ and $T$ as the energy flux and Unruh
temperature seen by an accelerated observer just inside the
horizon. Then Padmanabhan \cite{Padmanavan} was able to formulate
the first law of thermodynamics on the horizon, starting from
Einstein equations for a general static spherically symmetric
space time. The following nice equivalence
$$Laws ~of ~thermodynamics \Leftrightarrow Analogous~ laws ~of
~black ~hole~ dynamics
 ( Semi~ classical~ analysis) $$ $$~~~~~~~~~~~~~~~~\Leftrightarrow
 Einstein ~field ~equations (gravity~
theory)~(classical~ treatment)$$

perhaps shows the strongest evidence for a fundamental connection
between quantum physics and gravity.\\

Subsequently, this identity between Einstein equations and
thermodynamical laws has been applied in the cosmological context
considering universe as a thermodynamical system bounded by the
apparent  horizon ($R_{A}$). Using the Hawking temperature
$T_A=\frac{1}{2 \pi R_A}$ and Bekenstein entropy $S_A=\frac{\pi
R_A^2}G$ at the apparent horizon, the first law of thermodynamics
(on the apparent horizon) is shown to be equivalent to Friedmann
equations \cite{Cai1} and the generalized second law of
thermodynamics (GSLT) is obeyed at the horizon. Also in higher
dimensional space time the relation was established  for gravity
with Gauss-Bonnet term and for the Lovelock gravity theory
(\cite{Bamba},{\cite{
Akbar}, \cite{Lancoz}}).\\

But difficulty arises if we consider universe to be bounded by
event horizon. First of all, in the usual standard big bang model
cosmological event horizon does not exists. However the
cosmological event horizon separates from that of the apparent
horizon only for the accelerating phase of the universe (dominated
by dark energy). Further, Wang et.al. \cite{Wang} have shown that
both first and second law of thermodynamics break down at the
event horizon, considering the usual definition of temperature and
entropy as in the apparent horizon. According to them the
applicability of the first law of thermodynamics is restricted to
nearby states of local thermodynamic equilibrium while event
horizon reflects the global features of space time. Also due to
existence of the cosmological event horizon, the universe should
be non-static in nature and as a result the usual definition of
the thermodynamical quantities on the event horizon may not be as
simple as in the static space-time. They have considered the
universe bounded by the apparent horizon as a Bekenstein system as
Bekenstein's entropy-mass bound : $S \leq 2 \pi R_A$ and
entropy-area bound:$S \leq \frac{A}4$ are valid in this region.
These Bekenstein bounds are universal in nature and all
gravitationally stable special regions with weak self gravity
satisfy Bekenstein bounds. Finally, they have argued that as event
horizon is larger than the apparent horizon  so the universe
bounded by the event horizon is not a Bekenstein system.

In the literature, there are lot of works \cite{Wang1, Saridakis1,
Saridakis2, Charmouis, Kim, Gravanis, Cai2, Bousso},
 dealing with thermodynamics of the universe bounded by the
apparent horizon as it is a Bekenstein system. On the other hand,
due to the above complicated nature of the thermodynamical system
: universe bounded by the event horizon (UBEH), there are few
works related to it.  Recently, Mazumder et. al. \cite{Mazumder1,
Mazumder2}, starting from the first law of thermodynamics, have
examined the validity of the GSLT which states that the time
variation of the sum of the entropy of the horizon ($S_{H}$) and
the entropy of the matter inside it ($S_{I}$) should be positive
definite , i.e., $\frac{d}{dt}\left(S_{H}+S_{I}\right)\geq 0$.
Without assuming any specific choice for the entropy and the
temperature on the event horizon, they are able to show the
validity of the GSLT with some restrictions on the matter. In the
present work, we try to speculate the nature of the entropy
function on the event horizon assuming the validity of the GSLT.
It should be noted that the reason to stress the validity of GSLT
is that it is an universal law, holds in any generality,
irrespective of whether the thermodynamical system is an
equilibrium or in a non-equilibrium one.

We start with homogeneous and isotropic FRW model of the universe
having line element
\begin{equation}\label{1}
ds^{2}=h_{ab}dx^{a}dx^{b}+\tilde{}{r}^{2}d\Omega_{2}^{2}
\end{equation}
where $\tilde{r} = ar$ is the area radius, $h_{ab}=diag\left(-1,
\frac{a^{2}}{1-kr^{2}}\right)$ with $k=~0, ~\pm1$ for a flat,
closed and open model and $d\Omega_{2}^{2} =
d\theta^{2}+\sin^{2}\theta d\phi^{2}$ is the metric on unit
2-sphere. The Friedmann equations are
\begin{equation}\label{2}
H^{2}+\frac{k}{a^{2}}=\frac{8\pi G \rho}{3}
\end{equation}
and
\begin{equation}\label{3}
\dot{H}-\frac{k}{a^{2}}=-4\pi G \left(\rho+p\right)
\end{equation}
with energy conservation equation
\begin{equation}\label{4}
\dot{\rho}+3H\left(\rho+p\right)=0
\end{equation}
The apparent horizon, a null surface is characterized by
$$h^{ab}\partial_{a}\tilde{r}\partial_{b}\tilde{r}=0$$
and hence the radius of the apparent horizon has the expression
\begin{equation}\label{5}
R_{A}=\frac{1}{\sqrt{H^{2}+\frac{k}{a^{2}}}}
\end{equation}
On the other hand, the radius of the cosmological event horizon
(which exists for accelerating model of the universe in Einstein
gravity) is given by,
\begin{equation}\label{6}
R_{E}=a\int_{t}^{\infty}\frac{dt}{a}=a\int_{a}^{\infty}\frac{da}{Ha^{2}}
\end{equation}
As,
\begin{equation}\label{7}
R_{H}=\frac{1}{H}
\end{equation}
is the radius of the Hubble horizon, so depending on the curvature
the horizons are related by the following relations

$(I)~~ k=0~~  :   ~~~~$~
$$R_{A}=\frac{1}{H}=R_{H}<R_{E}$$

$(II)~~ k=-1~~  :   ~~~~$~
$$R_{H}<R_{A}<R_{E}$$

$(III)~~ k=+1~~  :   ~~~~$~
$$Either, ~~R_{A}<R_{E}<R_{H}$$
$$or, ~~R_{A}<R_{H}<R_{E}$$

In reference \cite{Mazumder1, Mazumder2}, Mazumder et. al. have
assumed the validity of the first law of thermodynamics at the
event horizon (i.e., the Clausius relation  $-dE=T_{E}dS_{E}$)
where the amount of energy crossing the event horizon during the
infinitesimal time $dt$ is given by
\begin{equation}\label{8}
-dE=4 \pi R_{E}^{3}H\left(\rho+p\right) dt
\end{equation}
Here the change in the horizon entropy during the small time $dt$
is
\begin{equation}\label{9}
dS_{E}=\frac{4\pi R_{E}^{3} H}{T_{E}}\left(p+\rho\right)dt
\end{equation}
where $\left(S_{E}, ~T_{E}\right)$ are the entropy and
temperature of the event horizon and $(\rho,~p)$ are energy
density and thermodynamic pressure of the fluid bounded by the
event horizon.

Now, to calculate the variation of the matter entropy  we shall
use the Gibb's equation (\cite{Izquierdo})
\begin{equation}\label{10}
T_{E}dS_{I}=dE_{I}+pdV
\end{equation}
where $S_{I}$ and $E_{I}$ are the entropy and the energy of the
matter distribution respectively. One may note that for
thermodynamical equilibrium, the temperature of the matter is
chosen as that of the event horizon (i.e., $T_{E}$). As
\begin{equation}\label{11}
V=\frac{4}{3}\pi R_{E}^{3}~~,~~E_{I}=\frac{4\pi}{3}\rho  R_{E}^{3}
\end{equation}
so from the Gibb's equation (\ref{10}) and using the energy
conservation relation (\ref{4}) the variation of matter entropy
has the expression
\begin{equation}\label{12}
dS_{I}=4\pi
R_{E}^{2}\left(\rho+p\right)\left(\dot{R}_{E}-HR_{E}\right)dt
\end{equation}
Differentiating the relation (\ref{6}) for $R_{E}$, we get
\begin{equation}\label{13}
\dot{R}_{E}=\left(H R_{E}-1\right)
\end{equation}
Hence combining equations (\ref{9}) and (\ref{12}) using
(\ref{13}), the time variation of the total entropy is given by
\begin{equation}\label{14}
\frac{d}{dt}\left(S_{E}+S_{I}\right)=4\pi
\left(\rho+p\right)\frac{R_{E}^{2}H}{T_{E}}\left(R_{E}-\frac{1}{H}\right)
\end{equation}
Based on the above calculation Mazumder et. al. \cite{Mazumder1,
Mazumder2} has obtained the following restrictions for validity of
GSLT (i.e., $\frac{d}{dt}\left(S_{E}+S_{I}\right)\geq 0$)

$(i)$ For flat and open FRW  universe the GSLT  is valid if the
weak energy condition $\rho+p>0$ is satisfied.\\

$(ii)$ For a closed model, the validity of the GSLT demands either
the weak energy condition is satisfied and
$R_{A}<\frac{1}{H}<R_{E}$ or the weak energy condition is violated
and $R_{A}<R_{E}<\frac{1}{H}$.\\

$(iii)$ For validity of GSLT, no specific form of entropy or
temperature on the event horizon is needed.\\

Here in this present paper, the thermodynamical study is rather in
a opposite way. We start with validity of GSLT and infer about
properties of the entropy function. In analogy with Bekenstein's
entropy-area relation, we assume the functional form of the
entropy at the event horizon be
\begin{equation}\label{15}
S_{E}=\frac{f(A)}{4G}
\end{equation}
with $A=4\pi R_{E}^{2}$, the area of the event horizon. So,
\begin{equation}\label{16}
\frac{dS_{E}}{dt}=\frac{f'(A)}{G}2\pi R_{E}\dot{R}_{E}
\end{equation}
where 'dash' denotes differentiation with respect to $'A'$. Thus
using the expression (\ref{12}) for variation of matter entropy,
the time variation of the total entropy is given by

\begin{equation}\label{17}
\frac{d}{dt}\left(S_{E}+S_{I}\right)=2\pi
R_{E}\left[\frac{\left(HR_{E}-1\right)}{G}f'(A)-\frac{2R_{E}}{T_{I}}\left(\rho+p\right)\right]
\end{equation}

where $T_{I}$ is the temperature of the matter distribution and we
have $T_{I}=T_{E}$ for equilibrium thermodynamics.\\

Now we shall examine the validity of GSLT in (a) quintessence era
and (b) phantom era.\\

(a)In quintessence era, weak energy condition ($\rho+p>0$)is
satisfied and so according to Davies \cite{Davies} $\dot{R_{E}}>0$
i.e. $R_{E}>\frac{1}{H}$ . So from equations (\ref{12}) (using
(\ref{13})) and (\ref{16}) horizon entropy will increase provided
$f'(A)>0$ while the matter entropy is decreasing with time. Thus
GSLT will be satisfied provided the expression within the square
bracket in equation (\ref{17}) is positive .\\

 (b) On the other hand, in phantom era, there is a violation of weak energy condition $(\rho+p<0)$ and
Sadjadi  \cite{Sadjadi} $\dot{R_{E}}<0$ i.e. $R_{E}<\frac{1}{H}$ .
So matter entropy will always increase and horizon entropy will
increase or decrease depending on $f'(A)<0$ or $f'(A)>0$ and as
before the expression within the square bracket in equation
(\ref{17}) should be positive for validity of GSLT.\\

Therefore, for the validity of GSLT, the entropy function f(A)
must have the following characteristic :

(a) In quintessence era :-
 f(A) is an increasing function of A ,i.e., $R_{E}$ such that
\begin{equation}\label{18}
f'(A)>\frac{2R_{E}G\left(\rho+p\right)}{T_{I}(HR_{E}-1)}
\end{equation}

(b)In phantom era: Either  f(A) is still an increasing function of
A with
\begin{equation}\label{19}
0<f'(A)<
\frac{2R_{E}G}{T_{I}}\left|\frac{\left(\rho+p\right)}{HR_{E}-1}\right|.
\end{equation}
 or $f(A)$ is decreasing function of A i.e. $f'(A)<0$ and GLST
is identically satisfied in that era.\\

To have smooth entropy function across the phantom barrier one
must have $f'(A)=0$ on the barrier. As a result both matter and
horizon entropy becomes constant on the phantom crossing. Thus the
entropy function has either of the two possible behavior:\\

I. The entropy function increases sharply in the quintessence era
so that the expression within the square bracket in equation
(\ref{17}) must have positive value to satisfy GLST and then
$f(A)$ reaches a maximum at the phantom crossing, subsequently
slowly decreases in the phantom era so that GSLT is satisfied
there. We speculate that the graph of the entropy function
throughout the quintessence and phantom era will be as in the
figure I with a maxima at phantom crossing.\\

II. The entropy function increases in both quintessence and
phantom era with a point of inflexion at the phantom crossing as
shown in FigII.\\

\begin{figure}

\includegraphics[height=1.5in]{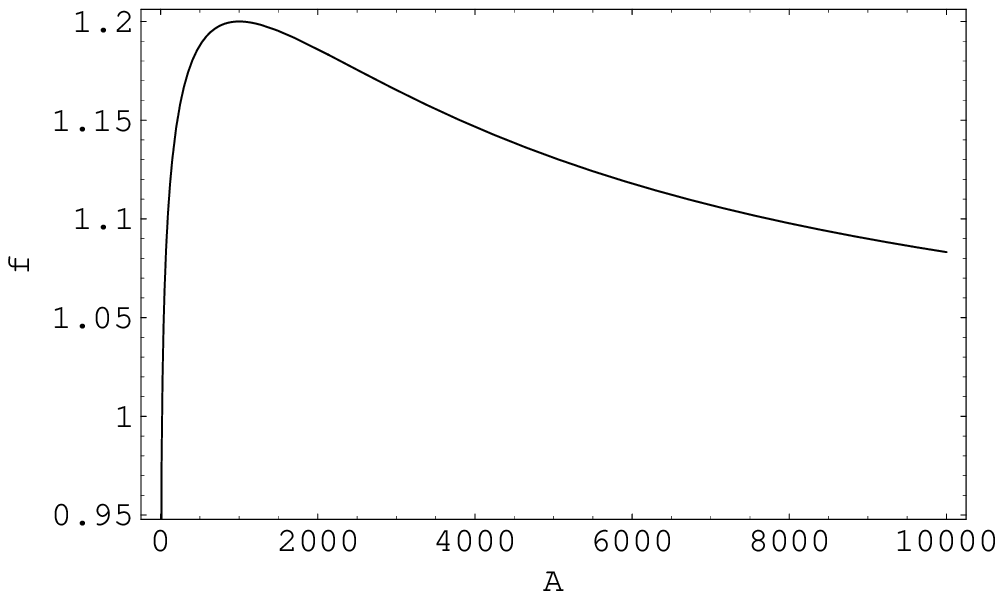}~~~
\includegraphics[height=1.5in]{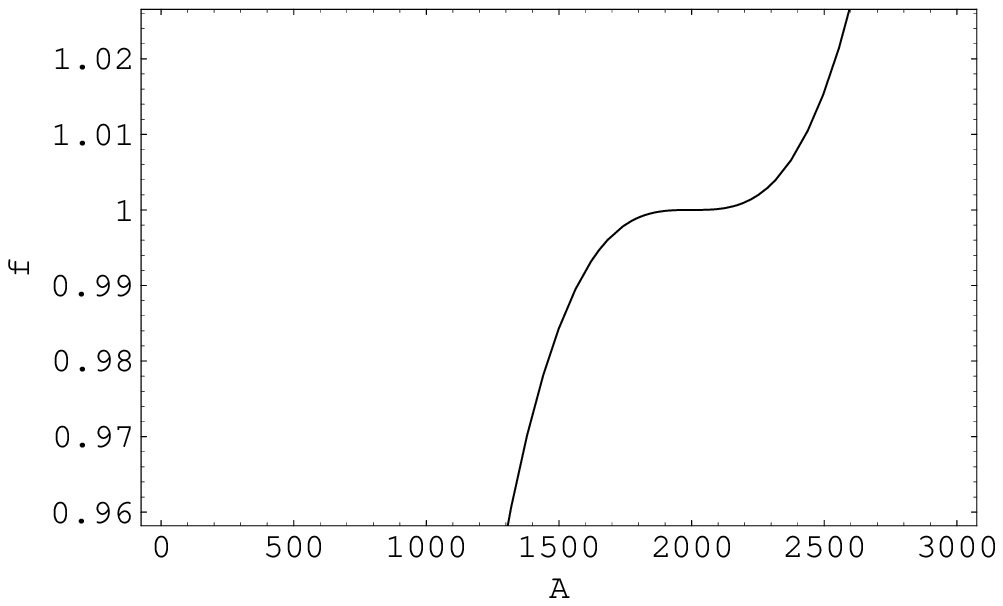}~~~

\vspace{1mm}
Fig.I~~~~~~~~~~~~~~~~~~~~~~~~~~~~~~~~~~~~~~~~~~~~~~~~Fig.II

\vspace{5mm} FigI. and  FigII.  shows the variation of the entropy function with the area
function \vspace{6mm}
\end{figure}

Now we shall examine whether the Bekenstein entropy-area relation
is valid on the event horizon. From the inequalities (\ref{18})
and (\ref{19}) to have GSLT on the event horizon with Bekenstein
entropy, the temperature of the matter distribution should satisfy
$$T_{I}>~{or}~< 2R_{E}G\left|\frac{(\rho+p )}{ HR_{E}-1}\right|$$ in
quintessence or phantom era respectively. This temperature bound
is quite distinct from Hawking temperature and we have
non-equilibrium thermodynamics. Finally, one may note that
throughout the calculation no specific Einstein field equations
have been used, only the equation of continuity is needed to
calculate the variation of matter entropy. Therefore, the above
result is true in any gravity theory.\\\\

{\bf Acknowledgement}
 RB wants to thank West Bengal State Govt,
India for awarding JRF. \\\\\\\\

\end{document}